# Measurement of viscoelastic properties of a liquid using an immersed rotating body of a general shape subjected to oscillatory shear

Hye Jin Ahn, Wook Ryol Hwang[a]

School of Mechanical Engineering, Gyeongsang National University, Jinju, 52828, Korea

## ABSTRACT

We propose a novel method for measuring linear and non-linear viscoelastic properties of a liquid by the oscillatory motion of an immersed rotating body in a vessel. The shape of a rotating object is general and we tested four different types of impellers: a disk, an anchor, and two different flat bladed turbines. In deriving the expressions of complex shear moduli, two different approaches were employed: one is based on the complex viscosity and the other is on the relationship between mean shear stress and mean shear strain. Both methods yield identical expressions for complex moduli. Using the latter method, the mean shear stress was appropriately scaled with torque, and the strain magnitude was scaled with the deflection angle, enabling its application to large-strain nonlinear oscillatory tests. Aqueous polyethylene oxide (PEO) solutions, xanthan gum solution and ketchup were tested and linear viscoelastic responses of storage and loss moduli were first presented as a function of the oscillation frequency. In spite of the presence of non-rheometric and highly non-uniform flow field, comparison with the data from the conventional cone-and-plate fixture of a rheometer shows remarkably accurate measurement with at most 7% average error within the frequency range from 0.01 [rad/s] to 100 [rad/s] for all the impeller geometries. In addition, large amplitude oscillatory shear experiments were also tested and discrepancy with highly elastic fluid were discussed. The proposed method may facilitate the in-situ measurement of viscoelastic properties of a fluid within an industrial reactor/agitator as a tool for on-line monitoring.

---

[a] The corresponding author. Tel.: +82-55-772-1628; Fax.: +82-55-772-1577; E-mail: wrhwang@gnu.ac.kr

## I. INTRODUCTION

Measurement of viscoelastic properties of a rheologically complex fluid is essential in understanding their internal structures and provides insights into how a material responds to deformation. It often provides an indirect window of observing the microstructural evolution within a material. For example, linear viscoelastic properties can be employed in the estimation of molecular weight distribution of polymers[1], particle size distribution within a polymer blend[2], and degree of curing in a reactive system[3]. While traditional off-line viscoelastic characterization using commercial laboratory rheometers is essential for this purpose, it is not sufficient for capturing the dynamic, time-sensitive changes that occur in industrial processes as materials experience flow, deformation and temperature variations, which is of great importance in optimal control of the process. In this context, real-time (in-situ or online) monitoring of the viscoelastic properties of a complex fluid is becoming increasingly necessary.

In this work, we present a novel viscoelastic measurement method utilizing complex-shaped impellers, which serves as a key foundational technology for the real-time viscoelastic monitoring in mixing processes. Since the mixing process with impellers is extensively employed during the production of materials, viscoelastic monitoring in mixing processes is of great importance across various material industries in setting up the proper mixing condition based on the microstructure. Despite this significance, previous studies on viscoelastic measurement techniques using complex rotating bodies, such as impellers, are not readily found. Instead, there are a large number of previous works in measuring the viscosity behavior as a function of shear rate for non-Newtonian fluids. The Metzner-Otto (MO) method[4] has been commonly employed for this purpose, which estimates the average shear rate by analyzing the total energy dissipation rate of non-Newtonian fluids and derives the corresponding viscosity based on torque and power measurements. Successful applications for the viscosity measurement were reported: e.g., for a helical ribbon impeller[5], an anchor impeller[6], and a flat bladed turbine[7].

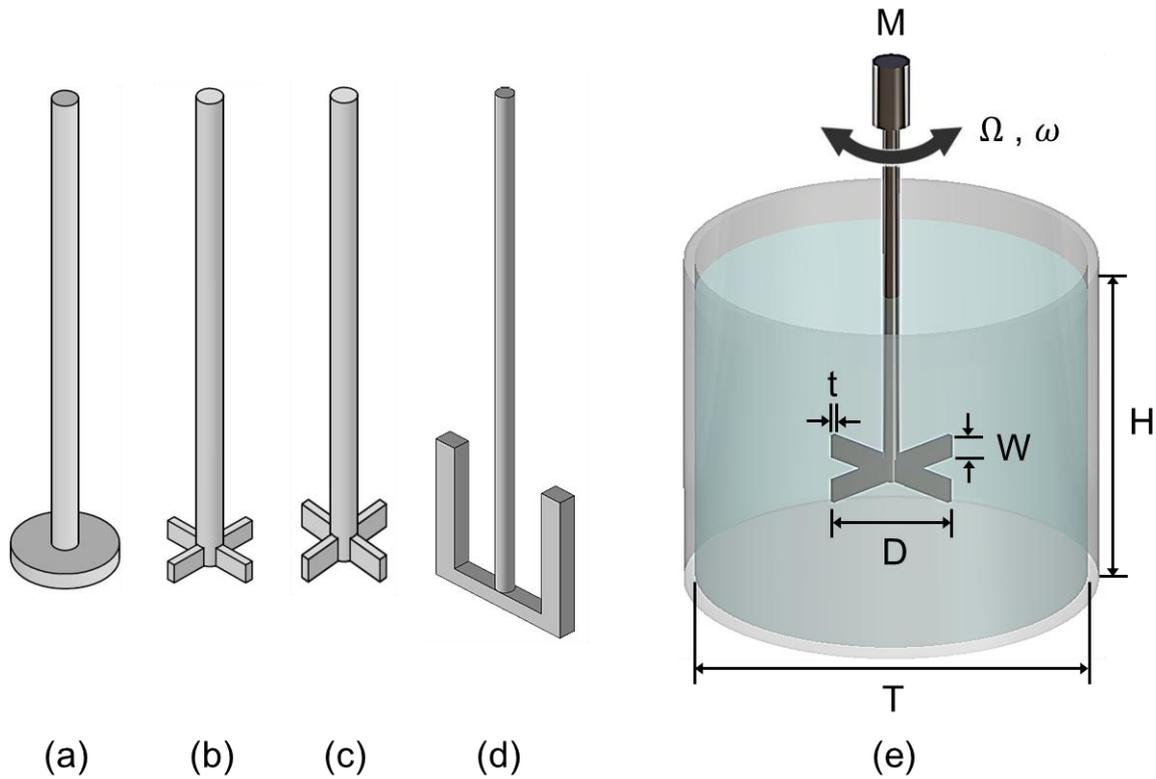

**Figure 1**: Geometries of rotating objects: (a) a disk; (b) a narrow flat-bladed turbine (FBTn); (c) a wide flat-bladed turbine (FBTw); (d) an anchor; (e) a vessel geometry with a rotating object.

We aim to propose a method to measure linear and non-linear viscoelastic properties of a liquid using the oscillatory motion of a rotating body with a general shape in a vessel. The system of interest is a typical agitator, consisting of a rotating body immersed in a liquid inside a vessel connected to a shaft, through which angular motion is controlled and the torque is measured. The goal of the study is to derive proper expressions of complex shear moduli directly from the relationship between angular rotation and torque subjected to oscillatory shear. In addition, one needs closed form expressions for the mean shear stress and the mean strain magnitude, which have to be determined by a torque and a deflection angle of the shaft, respectively, to extend this method for large angle oscillatory shear tests. The shear stress should be scaled correctly from the torque and the strain magnitude needs to be explicitly written by the deflection angle. These are challenging issues, as the flow field is non-rheometric and highly non-uniform and, moreover, the formation of flow field is highly dependent on rheological properties of a fluid and flow conditions. For example, the presence of unyielded region (cavern formation) is often observed with a yield stress fluid[8] and the reversal of secondary flow direction is well known phenomena for a disk rotating in a viscoelastic liquid due to the competition between inertia-driven and elasticity-driven secondary vortices of highly elastic fluids[9]. We will circumvent this complexity (i) by limiting the small strain amplitude for linear viscoelastic problems, (ii) by deriving the complex shear moduli through the complex viscosity, and (iii) by scaling the stress and strain magnitude properly. The last item of the stress and strain scaling is more general as it is valid even for non-linear large strain problems. The theoretically obtained results on viscoelastic measurement will be verified by experimental data with various geometries and fluids in both linear and non-linear regimes. We will report and discuss the feasibility and limitation of the proposed method.

## II. THEORETICAL DEVELOPMENT

### A. Determination of complex shear moduli from complex viscosity

*1. The relationship between torque and angular velocity or frequency*

We begin with the viscoelastic modulus identification in a small-strain linear viscoelastic regime using the complex viscosity. The problem is described in Fig.1(e) for an immersed rotating body in a vessel. The torque $M$ scales linearly with the angular velocity $\Omega$ (Fig. 1) in a Newtonian fluid with viscosity $\eta$ in a laminar regime: i.e., $M = \alpha\eta\Omega$, and a proportionality constant $\alpha$ will be a function of the geometry such as shapes of a rotating body and a vessel, and the liquid height, etc. (In this work, the angular velocity in steady rotation is denoted by $\Omega$ and the angular frequency in an oscillatory motion is denoted by $\omega$, to avoid confusion.) We remark that, in the theoretical derivation, the presence of free surface or the use of a circular vessel are not necessary and theoretical results derived here can be equally applied for systems with different types of walls: i.e., fully confined/open to free surface, or cylindrical/non-cylindrical, etc.

Several methods are available to find the proportionality constant, as it is a straightforward relationship. Among others, we adopt a conventional flow characterization method with the power number and the Reynolds number employed in the mixing community.[10] The reason for this choice is related to the determination of the scaling magnitude of stress and strain from the torque and the angular displacement, respectively, which will be discussed in Sec. II.B. The present work will employ the MO correlation[4] for a non-Newtonian fluid with the power number and the Reynolds number relationship for this purpose.

The power number is the dimensionless power draw in a rotating object system, which is the mean energy dissipation rate divided the characteristic energy dissipation rate in a turbulent regime. Under the steady angular velocity $\Omega$ with a Newtonian fluid viscosity $\eta$, the total power draw $P$ is the product of torque and angular velocity, i.e. $P = M\Omega$, which is identical to the total viscous energy dissipation rate $\int_V \eta \dot{\gamma}^2 dV$ in a laminar regime with $\dot{\gamma}$ and $V$ being the shear rate and the fluid volume, respectively. The power number $N_p$ and the Reynolds number $Re$ are conventionally defined as follows:

(1a,b) $\quad N_p = \frac{2\pi N M}{\rho N^3 D^5} = \frac{\pi^3 M}{4\rho \Omega^2 R^5}, \quad Re = \frac{\rho N D^2}{\eta} = \frac{2\rho \Omega R^2}{\pi \eta}$,

where $\rho$, $N$, $\Omega$, and $R$ are the density, the impeller speed in revolutions per second, the impeller angular velocity in radians per second, and the impeller radius, respectively. Two different forms were presented in Eq. (1): the former ones are conventional definitions in the mixing community with $N$ and the impeller diameter $D$, and the latter ones are with the angular velocity $\Omega = 2\pi N$ and the radius $R = D/2$. The latter combination will be employed in the present work for consistency, following the commonly accepted nomenclature in rheology. For a Newtonian fluid in a laminar regime, the power number appears reciprocal to the Reynolds number $N_p \sim Re^{-1}$ and the proportionality constant is denoted by $K_p$, which will be called as the energy dissipation rate coefficient in the present work: i.e.,

(2) $\quad N_p Re = K_p$.

$K_p$ is only a function of geometrical features of agitators (impeller, vessel and liquid height) and Eq. (2) is valid among geometrically similar systems, as it is written in a dimensionless form. Using Eqs. (1) and (2), the expression of torque $M$ as a function of viscosity $\eta$ and angular velocity $\Omega$ can be easily obtained as

(3) $\quad M = \alpha \cdot \eta\Omega = \frac{2R^3}{\pi^2} K_p \cdot \eta\Omega, \quad \alpha = \frac{2R^3}{\pi^2} K_p$

The proportionality constant $\alpha$ may be regarded as a shape factor that relates the torque to the product of viscosity and angular velocity. We remark that Eq. (3) will be valid for a rotating object of a general shape immersed in a liquid. There is no restriction on the shape of the rotating object, except for symmetrical geometries to avoid a static unbalance or dynamically- and hydrodynamically-induced vibrations, which can influence the accuracy of torque measurement. Static unbalance is caused by non-symmetric mass distribution; dynamic and hydrodynamic vibrations arise due to unsymmetrical distribution of the second moment of inertia and rotational drag, respectively.

## 2. Complex shear moduli from complex viscosity under small amplitude oscillatory shear

Let us consider the small strain oscillatory shear problem for linear viscoelastic properties. Assuming that the linearity in Eq. (3) holds for oscillatory shear at a small amplitude oscillation with the frequency $\omega$, one can simply replace the angular velocity $\Omega$ by the angular velocity $\dot{\theta}$ to get the torque and angular frequency relationship for a small amplitude oscillation such that $M = 2R^3/\pi^2 \cdot K_p \cdot \eta\dot{\theta}$. A sinusoidal response of the torque with a phase lag is expected with a viscoelastic fluid, from which the complex viscosity $\eta^*$ can be determined. Then, from the complex viscosity, the expressions for complex shear moduli $G'$ and $G''$ can be obtained as a function of torque, deflection angle, frequency, and geometrical parameters.

Let us denote the oscillation of a shaft using the deflection angle $\theta(t) = \theta_0 \cos(\omega t)$ and the angular frequency $\dot{\theta}(t) = -\theta_0\omega\sin(\omega t)$. Then one expects sinusoidal fluctuation in the torque response with a phase lag, i.e., $M(t) = M_0\cos(\omega t + \delta)$, as long as the oscillation amplitude $\theta_0$ is small. The relationship between $M(t)$ and $\dot{\theta}(t)$ with Eq. (3) yields the closed form expression for the complex viscosity $\eta^*$. To simplify the derivation procedure, the complex notation is employed as usual. Let $\theta(t) = Re\{\theta_0 e^{i\omega t}\}$, $\dot{\theta}(t) = Re\{i\theta_0\omega e^{i\omega t}\}$ and $M(t) = Re\{M_0 e^{i(\omega t+\delta)}\}$. Using Eq. (3) with the angular frequency $\omega$, the complex viscosity

$\eta^*$ can be obtained as follows:

(4) $\quad \eta^*(\omega) = \eta' - i\eta'' = \frac{\pi^2}{2K_p R^3} \frac{M_0}{\theta_0 \omega} (\sin\delta - i\cos\delta).$

From the identity in the linear viscoelastic properties of $G' = \omega\eta''$ and $G'' = \omega\eta'$, one can derive the expression of the storage modulus $G'$ and $G''$:

(5a,b) $\quad G'(\omega) = \omega\eta''(\omega) = \frac{\pi^2}{2K_p R^3} \frac{M_0}{\theta_0} \cos\delta, \quad G''(\omega) = \omega\eta'(\omega) = \frac{\pi^2}{2K_p R^3} \frac{M_0}{\theta_0} \sin\delta.$

For a general shape of an oscillating object with a known $K_p$, the storage and loss moduli of a liquid can be determined using Eqs. (5a) and (5b) by measuring the torque for a given angular frequency with the amplitude $\theta_0$. The immediate question that arises in Eq. (5) is how small the angular deflection of a shaft $\theta_0$ to ensure a linear response of the torque, which is the issue related to the scaling of strain magnitude and will be discussed in the next section.

## B. Determination of complex shear moduli from scaling of stress/strain magnitude and large amplitude oscillatory shear

Eq. (5) in the previous section is complete enough to determine the linear viscoelastic property for a rotating object with the torque $M$ and the angular frequency $\omega$. However, there remains a fundamental question on the determination of the stress and strain magnitudes to identify the strain limit for a linear regime and, moreover, to extend this method to non-linear measurements. There must be a representative (or mean) shear stress $\bar{\tau}$ that corresponds to the torque $M$ and a representative (mean) shear strain $\bar{\gamma}_0$, according to the angular deflection of a shaft $\theta_0$. Relationships between them are not straightforward, but one can at least assume linear relationships between stress and torque and between strain and deflection angle. In this regard, the scaling of stress and strain magnitude is the problem of determination of scaling factors $\alpha_\tau$ and $\alpha_\gamma$ such that $\bar{\tau} = \alpha_\tau M$ and $\bar{\gamma}_0 = \alpha_\gamma \theta_0$.

### 1. Determination of the strain magnitude

Determination of the strain scaling factor $\alpha_\gamma$ for a given oscillating object with a deflection angle $\theta_0$ is the problem of seeking a representative (or mean) strain $\bar{\gamma}_0$ in a complex flow field and it can be determined by integrating the scaling relation of the oscillatory shear rate. The shear rate scaling with the angular velocity $\Omega$ of the rotating body will be identical to that with the angular velocity $\dot{\theta}$, as the relationship between them is purely kinematic. In finding the shear rate magnitude of steadily rotating impellers, the MO relationship, which is well-known in the mixing community for more than three scores of years, can be applied. The MO relationship presents the mean shear rate $\bar{\dot{\gamma}}$, i.e. a representative shear rate in the agitator, scales linearly with the impeller speed $N$:

(6) $\quad \bar{\dot{\gamma}}(\Omega) = K_s N = \frac{K_s}{2\pi}\Omega,$

where a proportionality constant $K_s$ is the MO constant. Interestingly, the MO constant is regarded as only a (weak) function of the geometry of impeller and vessel with liquid height, nearly independent of fluid rheology for inelastic non-Newtonian fluids. Integration of Eq. (6) with respect to time, one gets the strain magnitude as a function of the deflection angle $\theta$ in steady shear flow as follows:

(7) $\quad \bar{\gamma}(\theta) = \frac{K_s}{2\pi}\theta$

Since the relationship between the strain and deflection angle is purely kinematical, Eq. (7) can also be employed with the oscillatory shear flow. For a oscillating shaft $\theta(t) = \theta_0 \cos(\omega t)$, the corresponding shear strain $\gamma(t) = \bar{\gamma}_0 \cos(\omega t)$ can be expressed as the magnitude of deflection angle $\theta_0$ and the scaling factor $\alpha_\gamma$.

(8) $\quad \bar{\gamma}_0 = \frac{K_s}{2\pi}\theta_0 = \alpha_\gamma \theta_0, \qquad \alpha_\gamma = \frac{K_s}{2\pi}.$

In addition, the corresponding magnitude of shear rate $\bar{\dot{\gamma}}_0$ is written as $\alpha_\gamma \theta_0 \omega$.

## 2. Determination of the stress magnitude

In a steady rotating flow in an agitator, there have been reported that the mean viscosity at the angular velocity $\Omega$ is identical to the material viscosity at the corresponding mean shear rate $\bar{\dot{\gamma}}(\Omega)$ in Eq. (7)[6,7]. Therefore, the mean stress $\bar{\tau}$ must be written as $\bar{\tau} = \eta(\Omega)\bar{\dot{\gamma}}(\Omega)$ with the mean viscosity and mean shear rate at the angular velocity $\Omega$. From the relationship between the torque and the angular velocity $\Omega$ in Eq. (3), the viscosity at the angular velocity is written in terms of torque and angular velocity:

$$\eta(\Omega) = \frac{\pi^2}{2R^3}\frac{1}{K_p} \cdot \frac{M}{\Omega}.$$

Then the mean shear stress can be obtained by multiplying the viscosity by the mean shear rate in Eq. (6) both at the angular velocity $\Omega$:

(9) $\quad \bar{\tau} = \eta(\Omega)\bar{\dot{\gamma}}(\Omega) = \frac{\pi}{4R^3}\frac{K_s}{K_p} \cdot M = \alpha_\tau M, \quad \alpha_\tau = \frac{\pi}{4R^3}\frac{K_s}{K_p}.$

The above equation presents the general relationship between torque and mean shear stress in a rotating object in a vessel. We remark that both the strain scale factor $\alpha_\gamma$ and the stress scale factor $\alpha_\tau$ are function of the two flow numbers $K_s$ and $K_p$ along with the geometric dimension $R$, which means that one can determine the mean shear stress and strain of an impeller system with the two flow numbers obtained a priori for a given geometry set.

### 3. Complex shear moduli subjected to large oscillatory shear

The stress scale factor $\alpha_\tau$ in Eq. (9) is also valid for oscillatory flow, as the relationship between torque and shear stress presents the torque equilibrium. Once the proper scaling of stress and strain is done, one can derive alternatively the complex shear moduli $G'$ and $G''$ from the torque and deflection angle and they have to be identical to those from the complex viscosity in Eq. (5). Using the strain scale factor $\alpha_\gamma$ and stress scale factor $\alpha_\tau$, the complex oscillatory strain $\bar{\gamma}^*$ and the responsive complex shear stress $\bar{\tau}^*$ can be written as

(10) $\quad \bar{\gamma}^* = \bar{\gamma}_0 Re\{e^{i\omega t}\} = \alpha_\gamma \theta_0 Re\{e^{i\omega t}\}, \quad \bar{\tau}^* = \bar{\tau}_0 Re\{e^{i(\omega t+\delta)}\} = \alpha_\tau M_0 Re\{e^{i(\omega t+\delta)}\}.$

Then, the complex shear moduli $G'$ and $G''$ are expressed as

(11a) $\quad G'(\omega, \gamma_0) = Re\left\{\frac{\bar{\tau}^*}{\bar{\gamma}^*}\right\} = \frac{\alpha_\tau}{\alpha_\gamma}\frac{M_0}{\theta_0} Re\{e^{i\delta}\} = \frac{\pi^2}{2K_p R^3}\frac{M_0}{\theta_0}\cos\delta,$

(11b) $\quad G''(\omega, \gamma_0) = Im\left\{\frac{\bar{\tau}^*}{\bar{\gamma}^*}\right\} = \frac{\alpha_\tau}{\alpha_\gamma}\frac{M_0}{\theta_0} Im\{e^{i\delta}\} = \frac{\pi^2}{2K_p R^3}\frac{M_0}{\theta_0}\sin\delta, \quad \text{with} \quad \gamma_0 = \frac{K_s}{2\pi}\theta_0.$

Complex shear moduli in Eq. (11) from the mean stress and mean strain are identical to those in Eq. (5) from the complex viscosity, which confirm the scaling of mean shear stress and strain magnitude as well as the complex moduli expression in Eq. (5). In addition, the presence of the strain magnitude facilitates viscoelastic measurement under large oscillatory shear, as the relationship between torque and mean shear stress and that between strain magnitude and deflection angle must be valid naturally even for large strain problems.

## III. EXPERIMENTAL DETAILS

### A. Materials and preparation

In this work, four different types of fluids were employed as working fluids. Highly viscous silicone oil (KF-96-10000cs, Shinetsu Co., Japan) was selected as a reference Newtonian fluid. Three different aqueous polymer solution of 2wt%, 4 wt% and 5wt% with Poly(ethylene oxide) (PEO, Mw=4,000,000, Sigma-Aldrich Chemical Co., USA), 3wt% aqueous xanthan gum (Sigma-Aldrich Chemical Co., USA) solution, a commercial food stuff (Tomato ketchup, Heinz Co., USA) were prepared for validation experiments for the proposed viscoelasticity measurement method. In the preparation of the PEO solution, PEO particles were dried under vacuum over 1 or 2 days at room temperature. A mixer with a flat bladed turbine impeller were employed for the dissolution process. Mixing was performed at 300 rpm for the first three hours, followed by a slow mixing at 100 rpm for the remaining 21 hours. The diameter of vessel was 70mm with a liquid volume of 200mL and the diameter of impeller was 35mm. The xanthan solutions were prepared exactly in the same way. In case of the PEO solution, the temperature was maintained at 40 °C for the first 16 hours and it was lowered to the room temperature for the remaining 8 hours to avoid evaporation of the solvent. As for the xanthan gum solution, the temperature was kept constant at room temperature throughout the mixing process. The solution was considered to be completely homogeneous, when the steady shear viscosity varies less than 1% over three days period.

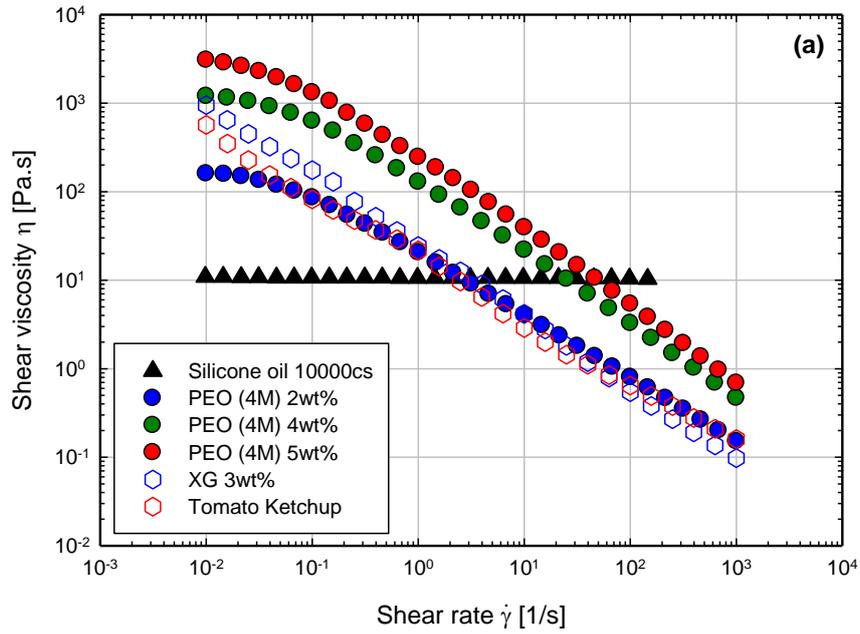

(a)

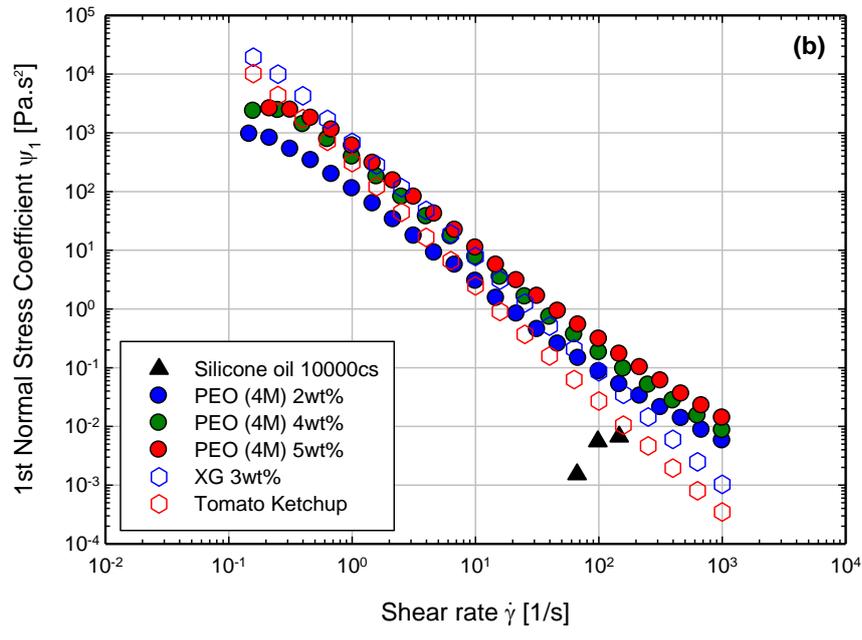

(b)

**Figure 2**: Rheological characterization of working fluids. (a) shear viscosity and (b) first normal stress difference as a function of shear rate.

The steady shear rheological behaviors were measured using a stress-controlled rheometer MCR301 (Anton Paar, Austria) equipped with a cone and plate (CP) fixture with a radius of 25 mm and a gap size of 0.05 mm. Plotted in Fig. 2 (a) and (b) are the shear viscosity and the first normal stress coefficient of all the working fluids as a function of the shear rate from 0.01 to 1000 1/s at room temperature. Working fluids are PEO solutions of 2wt%, 4wt% and 5 wt%, and xanthan gum solution (3wt%) and a tomato ketchup. The silicone oil shows inelastic Newtonian fluid behavior and all the others are viscoelastic shear-thinning fluids. In Fig. 2(a), the xanthan gum solution and tomato ketchup show yield behavior at low shear rate and three different PEO solutions exhibit zero-shear viscosity, which increases with the concentration and are approximately 200 Pa.s (2wt%), 1,500 Pa.s (4wt%) and 3,000 Pa.s (5wt%). Fig. 2(b) shows the first normal stress coefficients, which are all highly shear-thinning, except for the silicone oil. The first normal stress difference coefficients of PEO solutions increase with the concentration, as in the shear viscosity. For the shear rate ranges from 0.1 to 10, the ratio between first normal stress to shear stress, the Weissenberg number, appears $O(1)$, which confirms the elasticity of test fluids.

**Table 1.** Dimensional details of impellers.

|  | Disk | FBT4n | FBT4w | Anchor |
|---|---|---|---|---|
| Impeller diameter (D) [mm] | 12 | 12 | 12 | 26.02 |
| Impeller width (W) [mm] | 2.0 | 2.4 | 3.6 | 3.9 |
| Impeller thickness (t) [mm] | 1 | 1 | 1 | 1 |

**B. Rheometry with rotating objects**

To validate the proposed viscoelastic measurement scheme, four different types of impellers in a flat-bottomed vessel were built, as shown in Fig.1. The impellers were composed of a disk impeller (Fig. 1a), two flat bladed turbine impellers (Figs. 1b and 1c) and a close-clearance anchor impeller (Fig. 1d). Flat bladed turbine impellers were denoted by FBT4n and FBT4w, which are narrow- and wide-blade width impellers, respectively. Dimensional details can be found in Table 1. The impeller shaft was connected to the rheometer (MCR 301, Anton Paar, Austria) such that the torque and angular displacement measurement can be carried out accurately. The cup of a diameter 27 mm in the concentric cylinder measuring system (CC27,

Anton Paar, Austria) was employed as the vessel, where the impeller was submerged as indicated in Fig. 1(e). The liquid height was kept 30 mm for all the impeller systems in this work. Note that the torque value $M$ in this work is the measured torque with the rheometer, after compensating instrument inertia prior to sample measurements.

To investigate steady shear behaviors with impellers, steady rate-sweep measurements were firstly carried out at a fixed temperature of 20 °C over a wide range of shear rates from 0.01 to 1000 1/s with a logarithmically increasing scale. We remark that the shear rate in the impeller system was scaled with the MO constant $K_s$ (Eq. 6), which will be determined later, and that the scaled shear rate ranges roughly the same as with the standard CP fixture. Then to determine non-linear viscoelastic behaviors, dynamic oscillatory strain-sweep tests were performed over a strain amplitude range from 0.01 to 1000% with a logarithmically increasing scale at several fixed angular frequencies. Again the strain magnitude with the impeller system was scaled with the MO constant $K_s$ (Eq. 8) and it ranged similar to the strain amplitude of the CP fixture. Next dynamic frequency-sweep measurements were conducted to evaluate the linear viscoelastic behaviors in small amplitude oscillatory shear fields. Frequency-sweep experiments were tested over a broad range of angular frequencies from 0.1 to 100 rad/s with a logarithmically increasing scale at constant strain amplitudes of 0.1%. From preliminary strain-sweep tests, the strain magnitude of 0.1% was confirmed to lie within the linear viscoelastic region in both the standard CP and impeller geometries.

## IV. RESULTS AND DISCUSSION

### A. Determination of energy dissipation rate coefficient and MO constant

**Table 2.** Experimentally determined energy dissipation rate coefficient $K_p$ and MO constant $K_s$ for four different impeller geometries.

|  | Disk | FBT4n | FBT4w | Anchor |
|---|---|---|---|---|
| Energy dissipation rate coefficient $K_p$ | 101.5 | 68.21 | 81.50 | 321.3 |
| Effective shear rate constant $K_s$ | 13.39 | 10.46 | 11.10 | 32.53 |

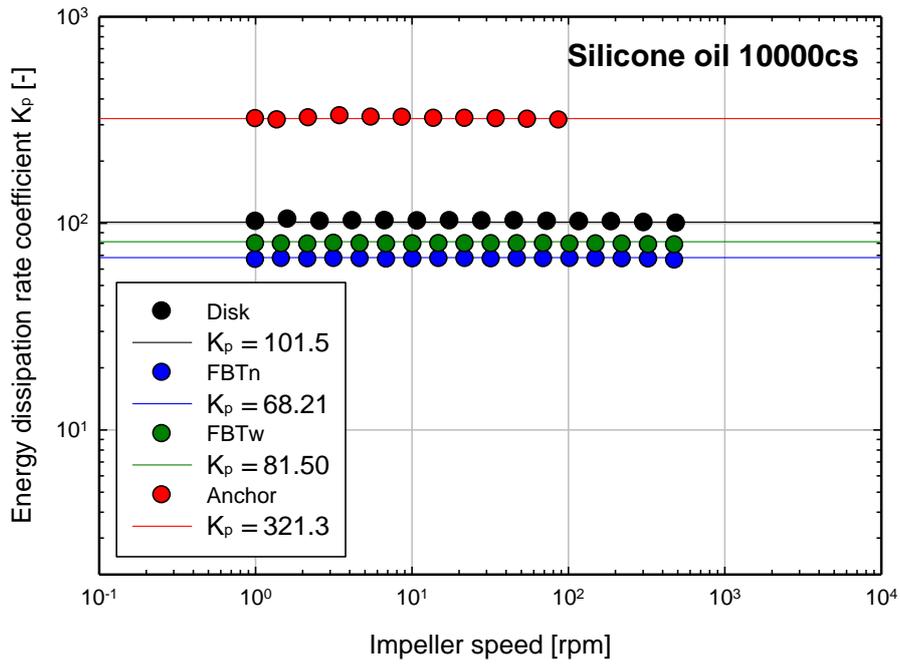

(a)

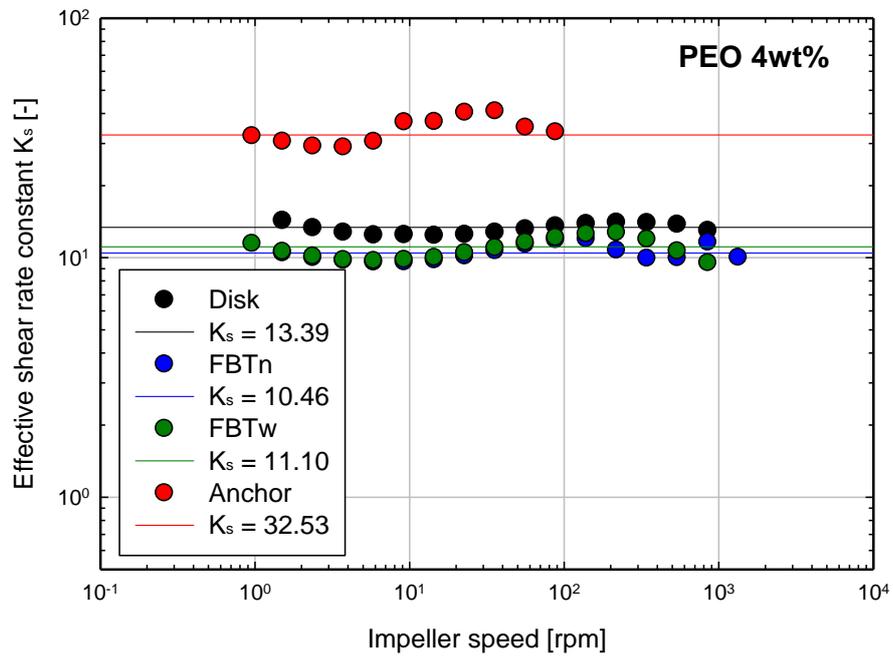

(b)

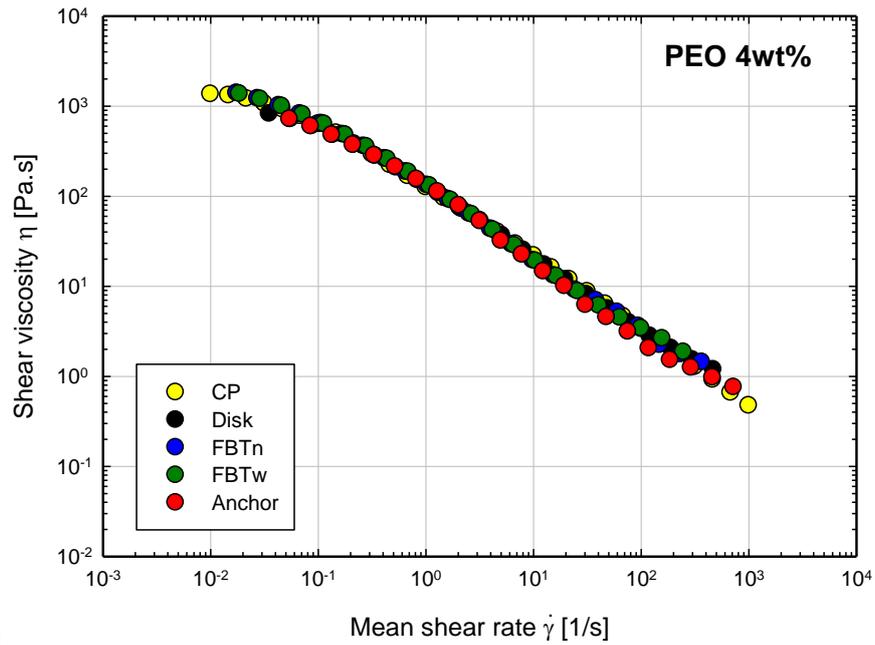

(c)

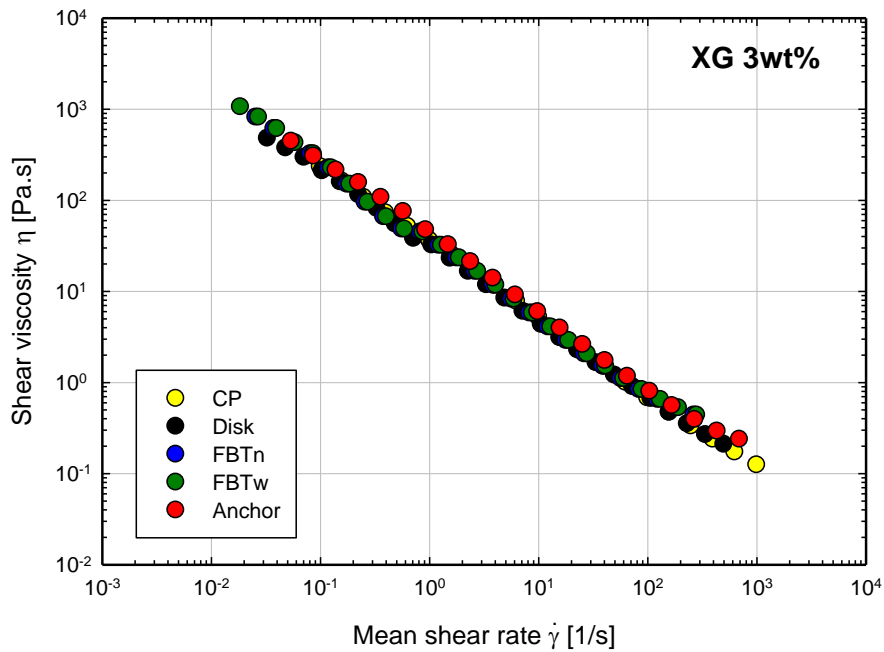

(d)

**Figure 3**: (a) The energy dissipation rate coefficients as a function of the impeller speed for disk, anchor, and two different flat disk turbines, measured with a Newtonian fluid (silicone oil); (b) The effective shear rate coefficients as a function of the impeller speed for disk, anchor, and two different flat disk turbines, measured with 4wt% PEO solution; Experimental validation with two flow constants with the shear viscosity estimation as a function of the effective shear rate in comparison with cone-and-plate measurement, for (c) 4wt% PEO solution and (d) 3wt% xanthan gum solution.

In order to identify viscoelastic properties from the torque response, one needs to find the energy dissipation rate coefficient $K_p$ (Eq. 2) and the MO constant $K_s$ (Eq. 8) a priori for a given impeller system, which is necessary to determine the magnitudes of stress and strain. From Eqs. (1) and (2), the determination of flow number $K_p$ can be accomplished by the product of the power number $N_p$ and the Reynolds number $Re$ for a given set of shaft torque $M$ and angular velocity $\Omega$ for a Newtonian fluid. In the present work, silicone oil (10,000 cs) was employed as a test fluid in determination of the torque and angular velocity relationship. Torque data of each impeller was measured for 30 different angular speeds from $N = 0.1$ [rpm] to 3300 [rpm]. Fig. 3(a) shows the energy dissipation rate coefficient $K_p$ as a function of the angular speed for all the four impellers and Table 2 lists the energy dissipation rate coefficient $K_p$ meausured in this way. The energy dissipation rate coefficient of disk, FBT4n and FBT4w were found $K_p = $ 101.5, 68.21 and 81.50, respectively, and that of the anchor impeller was 321.3. The corresponding maximum relative error in finding $K_p$ with Fig. 3(a) were 1.76% (disk), 1.66% (FBT4n), 1.43% (FBT4w), 2.71% (anchor) in experiments.

Next the MO constant $K_s$ was determined by assuming that the relationship between power number and Reynolds number is still valid even for non-Newtonian fluids, following the original MO correlation.[4] In this case, one can introduce a mean Reynolds number $\overline{Re}$, which is evaluated with the mean viscosity $\bar{\eta}$: i.e., $\overline{Re} = 2\rho\Omega R^2/\pi\bar{\eta}$. The power number and (mean) Reynolds number relationship is assumed to reproduce $N_p\overline{Re} = K_p$. The PEO solution (4wt%) was employed as a test fluid in determining the MO constant. The power number can be identified by measuring the torque for a given angular velocity $\Omega$ by Eq. (1) with non-Newtonian fluids and the mean Reynolds number and the viscosity are determined by $\overline{Re} = K_p/N_p$ and $\bar{\eta} = \frac{2\rho\Omega R^2}{\pi} \cdot \frac{N_p}{K_p}$ with the above determined energy dissipation coefficient $K_p$. Then the mean shear rate $\bar{\gamma}$ is identified by the corresponding shear rate of the mean viscosity using the fitted shear viscosity curve. The MO constant $K_s$ is then determined at each angular velocity by Eq. (6). The mean MO constant $K_s$ was obtained by averaging the MO constants. Presented in Fig. 3(b) are the MO constants as a function of the impeller speed and their average for each impeller geometry. The value of $K_s$ for each impeller geometry were 13.39 (disk), 10.46 (FBT4n), 11.10 (FBT4w), and 32.53 (anchor), respectively. The maximum relative error in $K_s$ were 7.61% (disk), 14.42% (FBT4n), 14.64% (FBT4w) and 25.75% (anchor).

Variation in the MO constant for various impeller speed does not seem to be negligible. However, the changes in $K_s$ does not significantly affect the viscosity or shear stress behavior, because of the dependence of shear rate is best shown in a log-log plot. Plotted in Figs. 3(c) and (d) are the shear viscosity as a function of the mean shear rate, measured with the four impeller geometries for two different fluids. The viscosity measurement with the impellers was made using Eq. (3) with a number of data set of the shaft torque $M$ and the angular speed $\Omega$ and the mean shear rate was scaled with the MO constant in Table 2 using Eq. (6). Viscosity behavior measured with the standard CP fixture was compared with that from the impellers for 4wt% PEO solution (Fig. 3c) and 3wt% xanthan gum solution (Fig. 3d). Note that the 4wt% PEO solution was a reference fluid that was employed in finding the MO constant and that 3wt% xanthan gum solution is a test fluid to verify the accuracy of the MO constant obtained with the PEO solution. In both cases, the shear viscosity from the impeller geometry with respect to the mean shear rate reproduces the materials shear viscosity and shear rate relationship reasonable well for a wide range of the shear rate, capturing the yield stress behaviors of the xanthan gum solution and zero-shear viscosity behavior of the PEO solution.

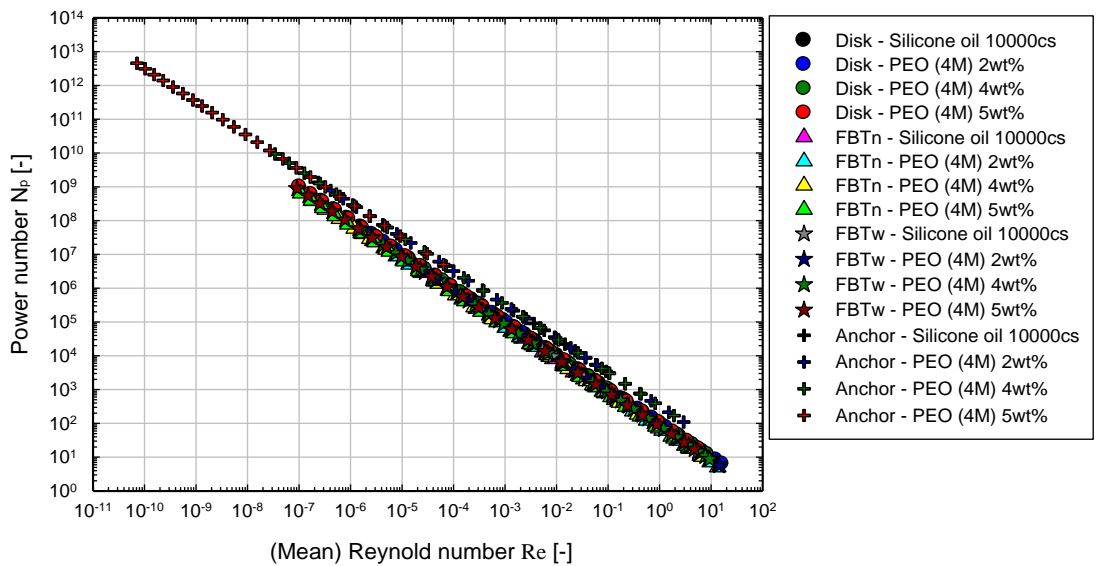

**Figure 4**: The relationship between power number and (mean) Reynolds number for all the four impeller geometries with a Newtonian fluid (silicone oil) and three different concentration of PEO solutions.

Once the energy dissipation rate coefficient $K_p$ and MO constant $K_s$ are determined correctly, the power number $N_p$ and mean Reynolds number $\overline{Re}$ relationship must be identical to that of a Newtonian fluid between $N_p$ and $Re$ for a given impeller system. In Fig. 4, the $N_p$ and $\overline{Re}$ relationship were plotted with PEO solutions with three different concentrations (2wt%, 4wt% and 5wt%), in comparison with the Newtonian relationship between $N_p$ and $Re$ with the test fluid (silicone oil). The $N_p - \overline{Re}$ relationships of non-Newtonian fluids with different PEO concentrations collapse into a single master curve for each geometry, which is identical to that of a Newtonian fluid. This result is natural, as the MO constant was determined to satisfy the $N_p - \overline{Re}$ to be identical as that of a Newtonian fluid, though averaged.

## B. Measurement of linear viscoelastic properties

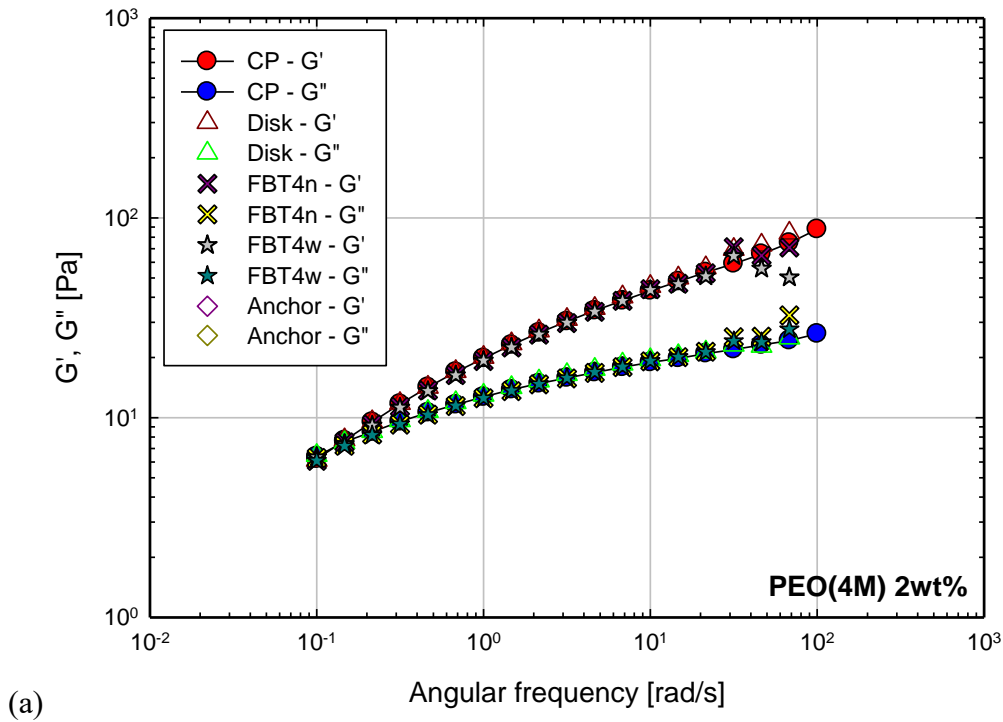

(a)

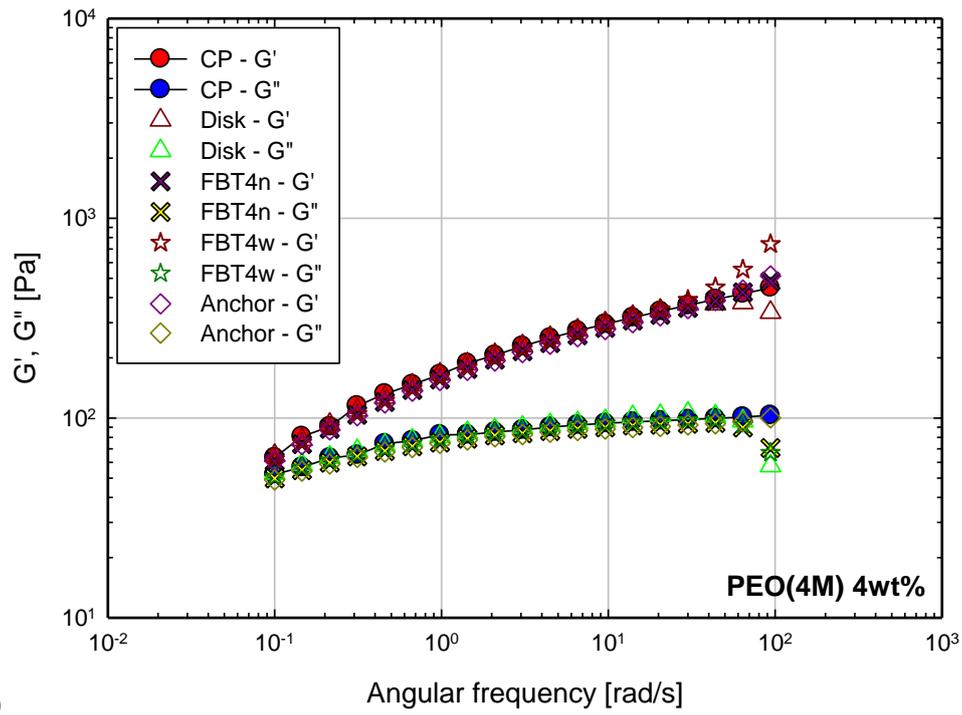

(b)

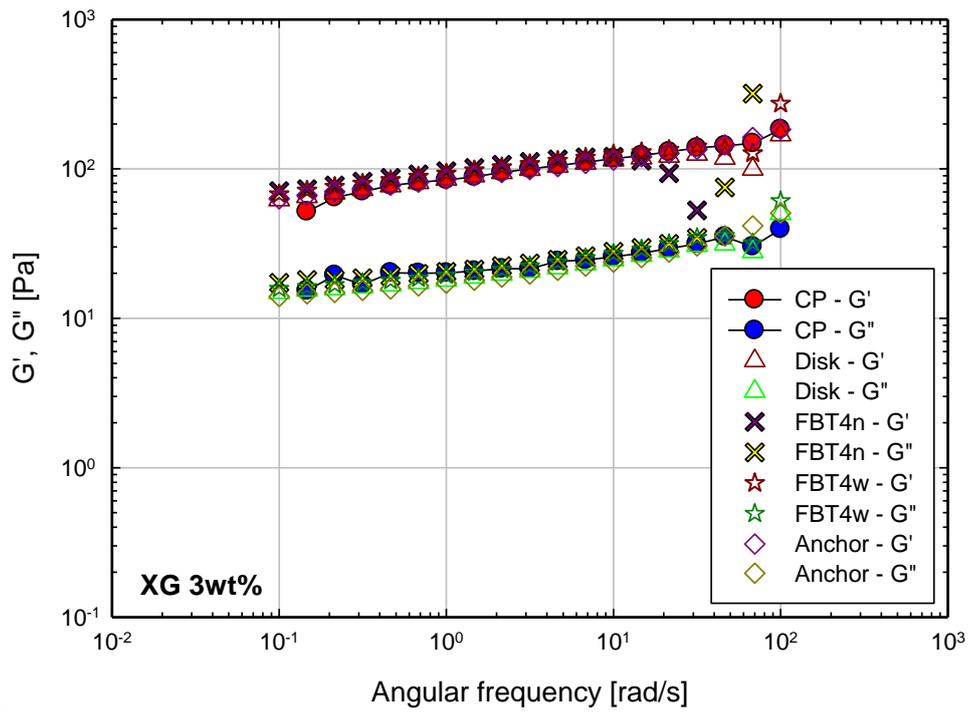

(c)

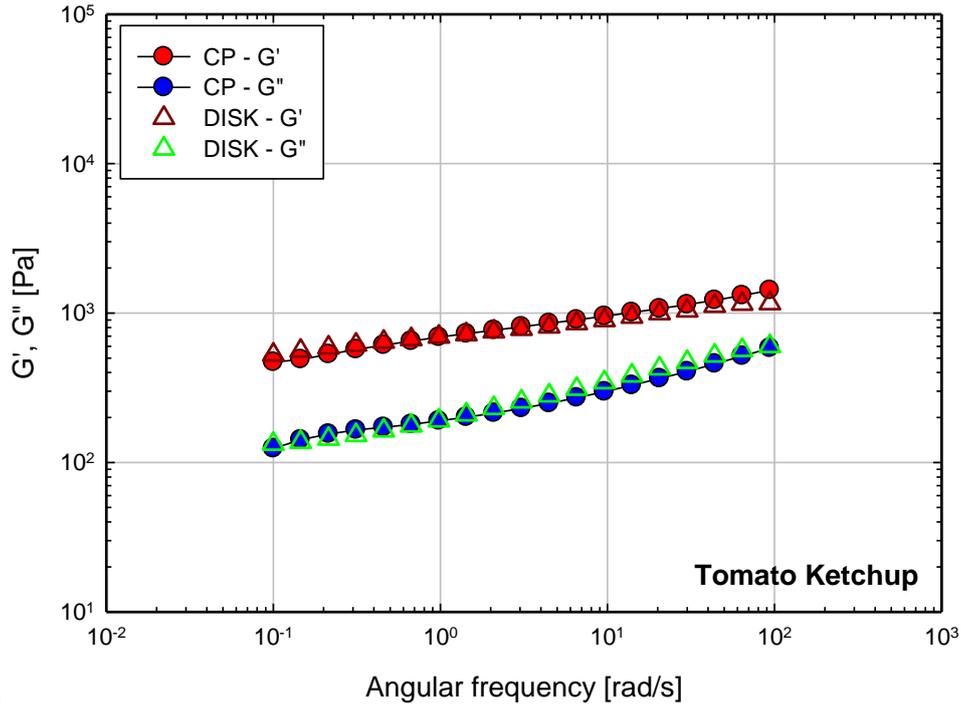

(d)

**Figure 5**: The linear viscoelastic properties of $G'$ and $G''$ as a function of the oscillation frequency $\omega$ measured with various impeller geometries, in comparison with the standard CP (cone-and-plate) data: (a) 2wt% PEO solution; (b) 4wt% PEO solution; (c) 3wt% xanthan gum solution; (d) tomato ketchup.

Firstly, we tested the proposed viscoelastic identification method for linear viscoelastic properties of $G'$ and $G''$ as a function of the oscillation frequency $\omega$. In the experiment, the shaft deflection angle was set by $\theta_0 = 6.28 \cdot 10^{-3}$ [rad] and the corresponding mean strain magnitude $\bar{\gamma}_0$ can be determined by Eq. (7), which indicates that the mean strain is determined by the deflection angle and the MO constant $K_s$ of the impeller. The actual values of the mean strain amplitude were 1.339 % (disk), 1.046% (FBT4n), 1.110% (FBT4w), and 3.253% (anchor), respectively. As will be shown later with the result of large oscillatory shear, the deflection angle $\theta_0$ in the present experiment is confirmed to assure the linear viscoelastic regime. $G'$ and $G''$ were computed by Eqs. 5(a) and (b), respectively, using the torque amplitude $M_0$, deflection angle $\theta_0$, energy dissipation rate coefficient $K_p$ and phase angle $\delta$.

Figs. 5(a-d) show linear viscoelastic properties of $G'$ and $G''$ as a function of the oscillation frequency $\omega$ for four different non-Newtonian fluids: PEO 2wt% (Fig. 5a), PEO

4wt% (Fig. 5b), xanthan gum 3wt% (Fig. 5c) and a tomato ketchup (Fig. 5d). The frequency ranges from 0.1 to 100 [rad/s]. In Fig. 5(a-c), measured $G'(\omega)$ and $G''(\omega)$ data with the four different impeller geometries (disk, FBT4n, FBT4w and anchor) were compared with the results from the standard CP fixture for 2wt% and 4wt% PEO solutions and 3wt% xanthan gum solution, respectively. In Fig. 5(d), the comparison was made for the disk with the CP fixture for the ketchup. As indicated in Fig. 5(a-d), $G'$ and $G''$ data with various impeller geometries were found surprisingly accurate, except for a high frequency larger than 20 [rad/s]. Roughly speaking, the data with flat bladed turbine deviates at a relatively lower angular frequency than those of disk and anchor, compared to the reference data from the CP fixture. To investigate the error quantitatively, the exact values of $G'$ and $G''$ and relative errors to the CP data were listed in Table A (supplementary data) with 4wt% PEO solution for all the impeller geometries. From Table A, the average relative errors in $G'$ were 3.47% (disk), 5.24% (FBT4n), 9.05% (FBT4w), and 6.67% (anchor) and those in $G''$ were 5.29% (disk), 6.57% (FBT4n), 3.69% (FBT4w), and 6.01% (anchor), which shows the accuracy in measuring linear viscoelatic properties using the impellers with complex geometries. Moreover, the investigation of $G'$ and $G''$ data at low frequeuncy (less than 10 [rad/s]) reveals that the maximum relative errors in both $G'$ and $G''$ are limited less than 3.8% (disk), 9.9% (FBT4n), 6.7% (FBT4w), and 11% (anchor). At a high frequency, the relative errors reach up to 44.19% (disk), 31.33% (FBT4n), 65.95% (FBT4w) and 14.78% (anchor). The error at a high frequency seems to be related with fluid inertia and elasticity and it will be discussed separately later.

**C. Measurement of non-linear viscoelastic properties: LAOS**

Plotted in Fig. 6 are the nonlinear viscoelastic data, including storage modulus $G'$ and loss modulus $G''$, obtained using four impeller geometries (Disk, FBT2, FBT3, and Anchor) with two different non-Newtonian fluids: 4 wt% PEO (4M) and 3wt% xanthan gum solutions, in both the linear regime, where the internal structure of the material remains intact, and the nonlinear regime subjected to large amplitude oscillatory shear (LAOS). Additionally, the nonlinear viscoelastic data measured using the viscoelastic measurement technique proposed in this study were compared with those from the cone-and-plate rheometry.

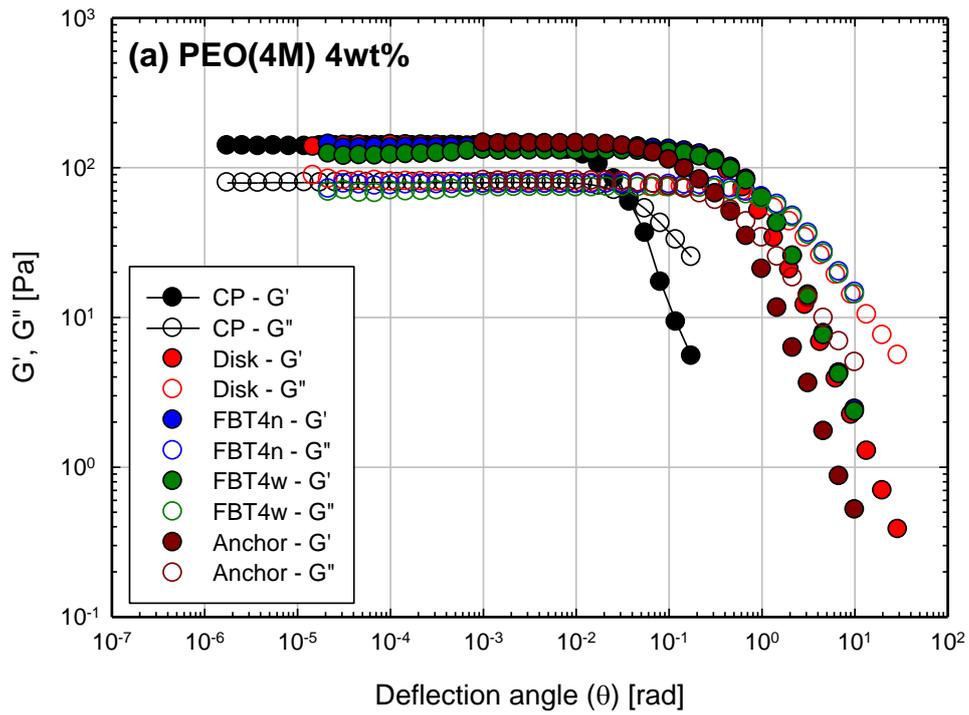
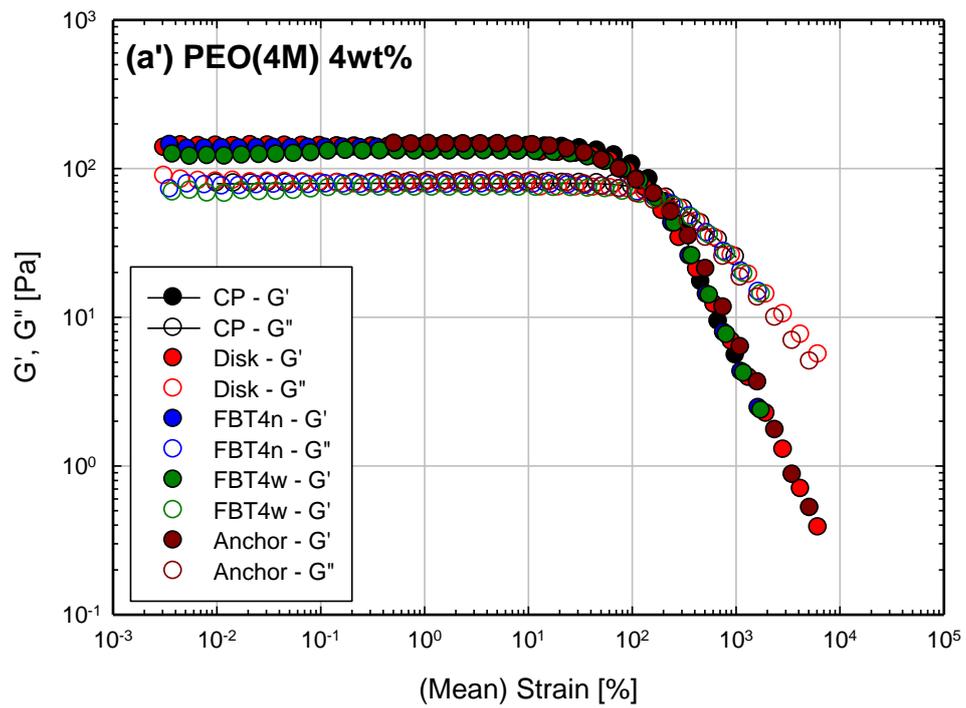

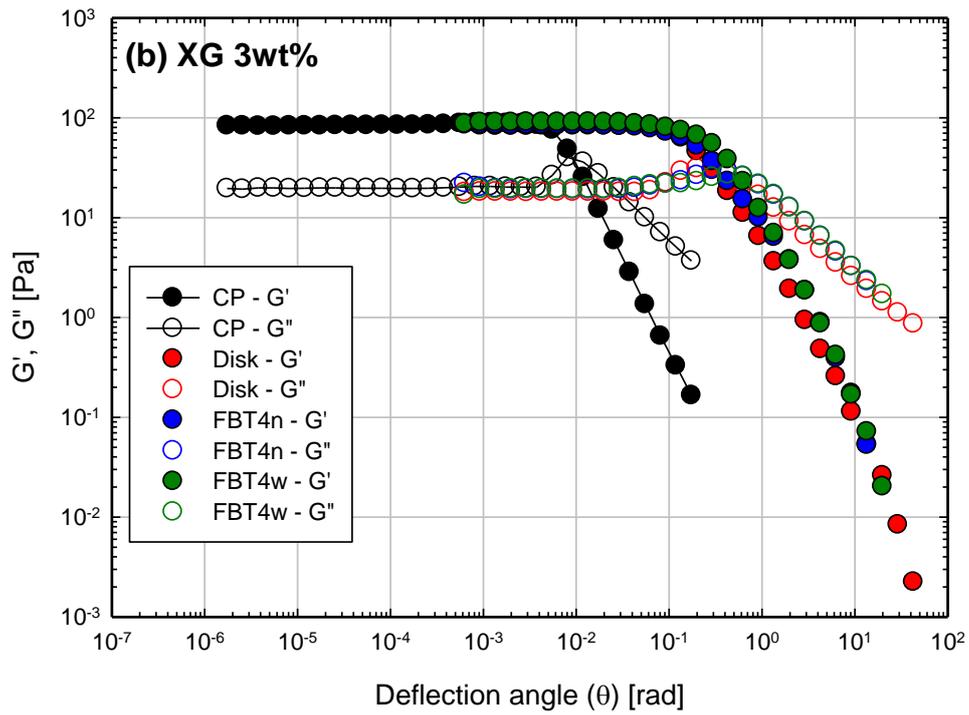

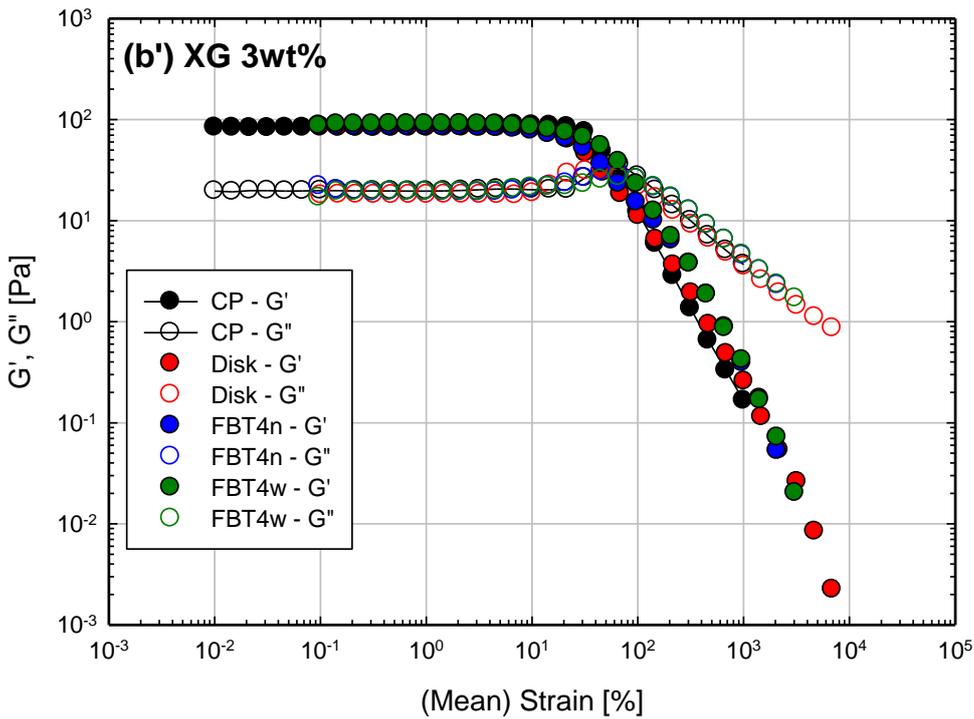

**Figure 6**: The non-linear viscoelastic properties of $G'$ and $G''$ as a function of either the deflection angle of the shaft (unprimed) or the corrected mean strain magnitude (primed) with four different impeller geometries, in comparison with the standard CP (cone-and-plate) data: (a) 4wt% PEO solution w.r.t. deflection angle; (a') 4wt% PEO solution w.r.t. strain magnitude; (b) 3wt% xanthan gum solution w.r.t. deflection angle; (b') 3wt% xanthan gum

solution w.r.t. strain magnitude. The angular frequency was set to 1 [rad/s] for all the data.

The measurement conditions for the nonlinear viscoelasticity were represented by deflection angles and the corresponding (corrected) mean strain magnitudes. The mean strain magnitude was computed by Eq. (8) with the MO constant listed in Table 2. For the cone-and-plate (CP) geometry, the deflection angle ranged from $1.75 \times 10^{-6}$ to $1.74 \times 10^{-1}$ [rad], with mean strain values between $1 \times 10^{-2}$ and $1.00 \times 10^{3}$ %. For the disk geometry, the deflection angle varied from $1.47 \times 10^{-5}$ to $2.92 \times 10^{1}$ [rad], and the mean strain spanned $4.31 \times 10^{-3}$ to $8.57 \times 10^{3}$ %. Similarly, for FBT2 and FBT3 geometries, the deflection angles ranged from $2.15 \times 10^{-5}$ to $1 \times 10^{1}$ [rad], with corrected strain magnitudes of $4.64 \times 10^{-3}$ to $2.16 \times 10^{3}$ % and $5.03 \times 10^{-3}$ to $2.34 \times 10^{3}$ %, respectively. The anchor geometry exhibited deflection angles from $1 \times 10^{-3}$ to $1 \times 10^{1}$ [rad], with corrected mean strain values between $5.73 \times 10^{-1}$ and $5.73 \times 10^{3}$ %. In all the experiments, the oscillation frequency was set to 1 [rad/s].

Figs. 6(a) and 6(b) present uncorrected data based on given deflection angles, revealing significant deviations in the measurement results among different impeller geometries. In contrast, Figs. 6(a') and 6(b') present the storage modulus and loss modulus values with respect to the corrected strain magnitude, demonstrating remarkable consistency across all the impeller geometries . As the strain amplitude (or deflection angle) increases, both the storage modulus and loss modulus exhibit linear behavior in the small-amplitude region (approximately below 10% for PEO solution and below 1% for xanthan gum solution), where the storage modulus remains constant. Under large deformations, the PEO solution exhibits strain-thinning nonlinear behavior with both the storage modulus and loss modulus decreasing. On the other hand, the XG solution exhibits weak strain overshoot behavior, where the storage modulus decreases steadily, while the loss modulus initially increases before subsequently reduction. The results plotted in Fig. 6 support that linear and nonlinear viscoelastic data, which have traditionally been measured exclusively using rheometric flow fields, can also be accurately measured in non-rheometric and non-uniform flow fields generated by impeller geometries.

## D. Discrepancy with a highly elastic liquid: effects of fluid inertia

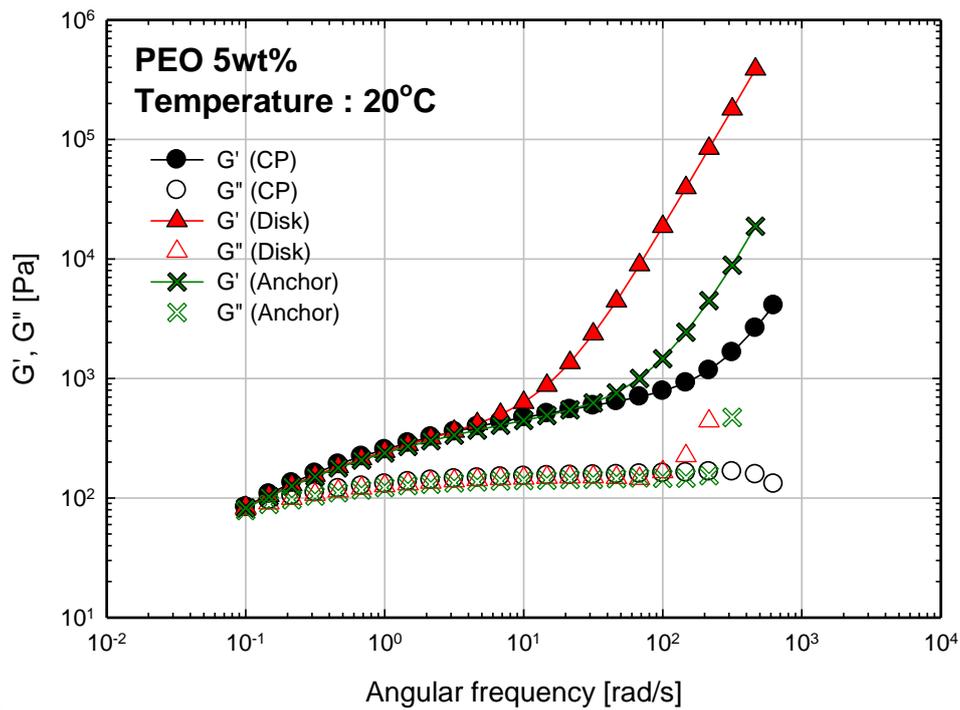

(a)

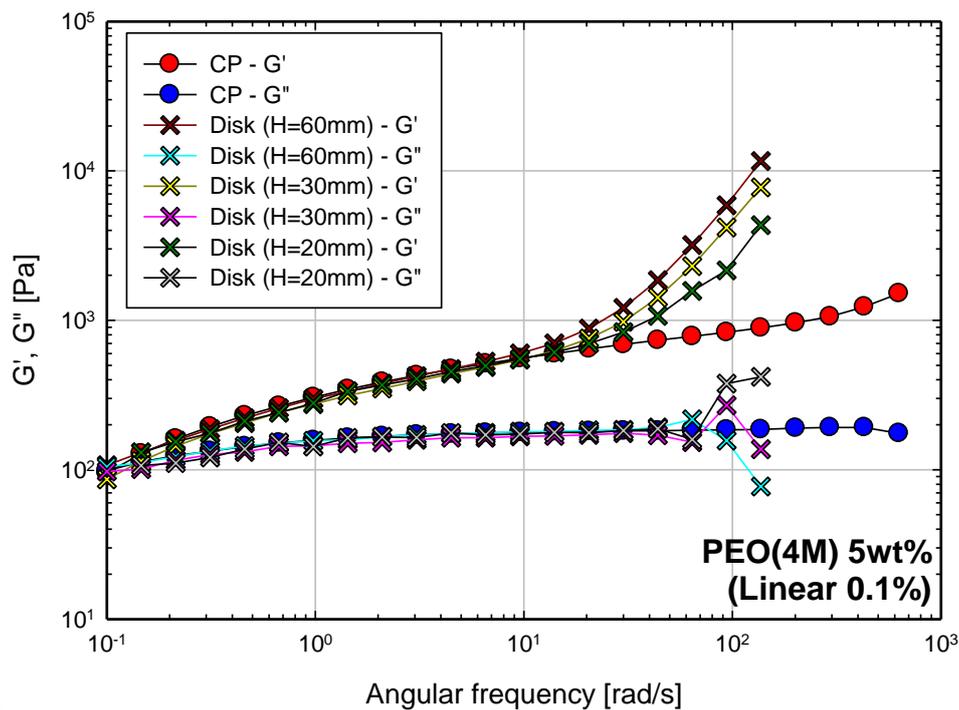

(b)

**Figure 7**: The storage modulus $G'$ and loss modulus $G''$ for a 5wt% PEO solution as a function of the frequency. (a) For disk and anchor impeller geometries and (b) disk geometries with three different liquid heights. Comparison was made with the cone-plate geometry.

As for the last topic, we will report the failure in the viscoelastic identification with the rotating objects. Fig. 7(a) shows the storage modulus $G'$ and loss modulus $G''$ with disk and anchor impeller geometries for a highly elastic 5wt% PEO solution in a linear regime with the strain magnitude 0.1% along with the data from the CP geometry for comparison. As indicated in Fig. 7(a), the storage moduli for the two geometries deviate from the standard CP data at moderate or high frequency, with disk impeller exhibiting larger deviation at lower frequencies compared to the anchor impeller. This discrepancy arises due to the effect of fluid inertia and is attributed to a large liquid volume involved in impeller-based rheometry. In general, there are two distinctive origins of inertia in rheological measurements with a combined mortor-transducer (CMT) rheometer: one is the instrument inertia, which is usually compensated with a modern commerical CMT rheometer, and the other is fluid inertia. The latter issue was recently studied by the author's group[11], where effects of fluid inertia on complex moduli was investigated for the parallel-plate geometry using the effective mechanical system analysis. In case of the parallel-plate (PP) geometry, deviation increases with the second moment of inertia of the liquid and, more specifically, it scales with the square of liquid height and frequency: i.e., $H_{PP}^2 \omega^2$, where $H_{PP}$ represents the gap thickness in a PP fixture. To systematically examine the effect of liquid volume, we tested the disk impeller with three different liquid heights, as shown in Fig. 7(b). As the liquid height increases, the deviation from the CP data becomes more pronounced and occurs at lower frequencies, as shown in Fig. 7(b). As the liquid height increases, the deviation from the CP data becomes more pronounced and occurs at lower frequences. The results in Fig. 7(b) are consistent with those reported by Gao and Hwang[11], indicating that fluid ineria signifantly impact the storage modulus at medium and hight frequencies, while the loss modulus remains unaffected.

## IV. CONCLUSION

In this work, we propose a systematic approach for measuring linear/non-linear viscoelastic properties of a liquid by the oscillatory motion of an immersed rotating body in a vessel. As for the rotating object, we tested the four geometries: a disk impeller, two flat bladed turbine impellers and a close-clearance anchor impeller, which are immersed in a liquid inside a cylindrical vessel. The expressions of complex shear moduli were derived in a closed form as a function of the impeller system geometry, the torque and the deflection angle, using the two methods: by converting complex viscosity into complex moduli and (ii) by scaling shear stress and strain based on torque and shaft deflection data. Both methods yield identical expressions for complex moduli. Notably, the expression for mean shear strain magnitude enables the measurement of nonlinear viscoelastic properties under large deformations. A comparison with SAOS data obtained from conventional CP fixture in a rheometer demonstrates highly accurate measurements, with an average deviation of at most 7% across all impeller geometries. Additionally, this method has been applied to LAOS experiments to assess nonlinear viscoelastic properties. The results on complex moduli in this study demonstrate that both linear and nonlinear viscoelastic data can be accurately measured using non-rheometric flow fields generated by impeller geometries.

The proposed method may provide a versatile approach for viscoelastic measurement, addressing challenges associated with conventional rheometry such as the wall slip, free surface solidification (or evaporation), and the edge fracture without a significant loss of accuracy [12]. Furthermore, as discussed previously, the measurement technique in this study has the potential to serve as a key technology for real-time viscoelastic monitoring in mixing processes, providing an effective tool for in-situ and online monitoring of microstructures. Real-time rheological measurements enable continuous process control through immediate quality assessment. As manufacturing progresses toward automation and smart factories, real-time rheological data can be seamlessly integrated into automated control systems, enabling predictive process management.

## Acknowledgments

The authors acknowledge financial supports from the National Research Foundation of Korea (NRF-2019R1A2C1003974).

## AUTHOR DECLARATIONS
### Conflict of Interest

The authors have no conflicts to disclose.

## Author Contributions

**Hye Jin Ahn**: Formal analysis (equal); Investigation (equal); Methodology (equal); Validation (equal); Writing − original draft (equal). **Wook Ryol Hwang**: Conceptualization (equal); Funding acquisition (equal); Investigation (equal); Methodology (equal); Project administration (equal); Supervision (equal); Writing – review & editing (equal).

## DATA AVAILABILITY

The data that support the findings of this study are available from the corresponding author upon reasonable request.

## SUPPLEMENTARY MATERIALS

Table A is provided in a separate file for the errors in measurement of linear viscoelastic data of 4wt% PEO solution with four impeller geometries in Fig. 5(b) in comparison to the standard CP data.

# Supplementary Material

**Table A.** Errors in measurement of linear viscoelastic data of 4wt% PEO solution with four impeller geometries in Fig. 5(b) in comparison to the standard CP data.

|  | CP | | Disk | | | | FBT4n | | | | FBT4w | | | | Anchor | | | |
|---|---|---|---|---|---|---|---|---|---|---|---|---|---|---|---|---|---|---|
| Frequency | G' | G" | G' | E(G') | G" | E(G") | G' | E(G') | G" | E(G") | G' | E(G') | G" | E(G") | G' | E(G') | G" | E(G") |
| [rad/s] | [Pa] | [Pa] | [Pa] | [%] | [Pa] | [%] | [Pa] | [%] | [Pa] | [%] | [Pa] | [%] | [Pa] | [%] | [Pa] | [%] | [Pa] | [%] |
| 0.100 | 63.30 | 52.20 | 64.76 | 2.311 | 52.81 | 1.174 | 60.60 | 4.266 | 49.78 | 4.634 | 62.23 | 1.698 | 50.89 | 2.511 | 61.11 | 3.460 | 48.97 | 6.195 |
| 0.146 | 80.80 | 56.60 | 77.80 | 3.711 | 58.45 | 3.266 | 73.99 | 8.431 | 54.75 | 3.260 | 75.45 | 6.617 | 56.59 | 0.017 | 73.58 | 8.936 | 54.20 | 4.236 |
| 0.214 | 90.50 | 63.10 | 92.71 | 2.446 | 63.49 | 0.626 | 88.25 | 2.491 | 59.68 | 5.416 | 89.67 | 0.919 | 61.41 | 2.679 | 87.24 | 3.608 | 58.82 | 6.778 |
| 0.313 | 115.0 | 65.20 | 109.2 | 5.026 | 68.60 | 5.216 | 103.7 | 9.864 | 64.14 | 1.619 | 105.0 | 8.701 | 65.82 | 0.955 | 102.6 | 10.76 | 63.43 | 2.713 |
| 0.458 | 132.0 | 74.30 | 126.8 | 3.948 | 72.97 | 1.786 | 120.2 | 8.923 | 68.34 | 8.019 | 122.3 | 7.371 | 70.23 | 5.476 | 118.9 | 9.888 | 67.54 | 9.098 |
| 0.669 | 147.0 | 76.90 | 145.8 | 0.795 | 77.21 | 0.397 | 138.0 | 6.117 | 72.12 | 6.217 | 140.5 | 4.389 | 74.41 | 3.240 | 136.1 | 7.410 | 71.13 | 7.509 |
| 0.979 | 165.0 | 82.30 | 165.5 | 0.284 | 80.62 | 2.044 | 155.9 | 5.534 | 75.34 | 8.461 | 159.7 | 3.204 | 78.08 | 5.130 | 154.2 | 6.571 | 74.51 | 9.466 |
| 1.430 | 188.0 | 82.60 | 186.1 | 1.036 | 84.14 | 1.867 | 175.7 | 6.541 | 78.80 | 4.604 | 179.2 | 4.675 | 81.24 | 1.641 | 173.2 | 7.865 | 77.60 | 6.059 |
| 2.090 | 206.0 | 85.00 | 207.0 | 0.487 | 87.21 | 2.596 | 195.8 | 4.954 | 81.67 | 3.918 | 200.4 | 2.707 | 84.34 | 0.773 | 192.8 | 6.391 | 80.44 | 5.370 |
| 3.060 | 229.0 | 87.20 | 229.0 | 0.016 | 89.79 | 2.975 | 215.9 | 5.742 | 83.84 | 3.855 | 221.6 | 3.225 | 86.88 | 0.362 | 212.6 | 7.156 | 82.58 | 5.297 |
| 4.480 | 251.0 | 89.90 | 252.0 | 0.388 | 92.31 | 2.681 | 237.0 | 5.590 | 85.96 | 4.378 | 243.8 | 2.884 | 89.30 | 0.666 | 233.9 | 6.800 | 84.86 | 5.603 |
| 6.550 | 273.0 | 92.30 | 275.0 | 0.725 | 94.47 | 2.351 | 259.1 | 5.087 | 88.10 | 4.546 | 267.7 | 1.952 | 91.68 | 0.676 | 255.3 | 6.479 | 86.54 | 6.238 |
| 9.580 | 295.0 | 94.10 | 298.4 | 1.137 | 96.63 | 2.687 | 282.2 | 4.325 | 89.74 | 4.637 | 292.5 | 0.861 | 93.70 | 0.428 | 276.7 | 6.205 | 87.97 | 6.511 |
| 14.00 | 318.0 | 95.60 | 322.0 | 1.250 | 101.3 | 5.910 | 307.3 | 3.370 | 91.74 | 4.042 | 320.1 | 0.666 | 95.57 | 0.034 | 299.9 | 5.694 | 89.53 | 6.350 |
| 20.50 | 342.0 | 96.80 | 345.4 | 1.001 | 102.7 | 6.048 | 328.3 | 4.011 | 92.48 | 4.464 | 349.8 | 2.290 | 96.53 | 0.278 | 324.4 | 5.157 | 90.61 | 6.399 |
| 30.00 | 366.0 | 98.10 | 370.8 | 1.301 | 105.6 | 7.647 | 355.8 | 2.800 | 93.45 | 4.745 | 390.3 | 6.652 | 97.79 | 0.318 | 352.4 | 3.711 | 92.20 | 6.009 |
| 43.90 | 393.0 | 99.60 | 371.5 | 5.463 | 102.1 | 2.543 | 385.1 | 2.014 | 94.39 | 5.233 | 451.1 | 14.78 | 97.88 | 1.726 | 387.1 | 1.513 | 93.58 | 6.046 |
| 64.10 | 418.0 | 101.0 | 376.9 | 9.821 | 96.47 | 4.481 | 425.4 | 1.765 | 89.51 | 11.38 | 553.6 | 32.43 | 92.63 | 8.284 | 436.3 | 4.371 | 95.56 | 5.386 |
| 93.80 | 448.0 | 103.0 | 337.4 | 24.69 | 57.49 | 44.19 | 482.4 | 7.685 | 70.73 | 31.33 | 743.5 | 65.95 | 67.11 | 34.85 | 514.2 | 14.78 | 100.0 | 2.896 |
| Average Error | | | | 3.465 | | 5.289 | | 5.237 | | 6.566 | | 9.051 | | 3.686 | | 6.672 | | 6.008 |